\documentclass[11pt]{article}
\usepackage[english]{babel}
\usepackage[latin1]{inputenc}
\usepackage{amsmath}
\usepackage{hhline}
\usepackage{amssymb}
\usepackage{latexsym}
\usepackage{xspace}
\usepackage{fancybox}
\usepackage{multicol}
\usepackage{pstricks}
\usepackage{pst-node}
\newcommand{\ignore}[1]{}

\newcommand{\IC}{{\it IC}}

\newcommand{\cL}{{\cal L}}

\newcommand{\rpr}[2]{{\it Repairs\/}_{#1}({#2})}

\newtheorem{definition}{Definition}
\newtheorem{theorem}{Theorem}
\newtheorem{corollary}{Corollary}
\newtheorem{proposition}{Proposition}
\newtheorem{lemma}{Lemma}

\newtheorem{example}{Example}

\newenvironment{proof}{\noindent{\em Proof.}}{\hfill\eop}

\title{\bf Minimal-Change Integrity Maintenance Using Tuple Deletions\thanks{Research supported by
NSF Grant IIS-0119186.}}

\author{{\bf Jan Chomicki}\thanks{Contact author. Address: Dept. CSE,
201 Bell Hall, Univ. at Buffalo, Buffalo, NY 14260-2000. Fax: (716) 645-3464.
Phone: (716) 645-3180, ext.103.}\\
University at Buffalo\\
Dept. CSE\\
chomicki@cse.buffalo.edu\\
\and {\bf Jerzy Marcinkowski}\\
Wroclaw University\\
Instytut Informatyki\\
jma@ii.uni.wroc.pl}
 
\newcommand{\eop}{\hbox{\hskip6pt\vrule height 6pt width 6pt}}
\begin{document}
\maketitle
\begin{abstract}
We address the problem of minimal-change integrity maintenance in the
context of integrity constraints in relational databases.  We assume
that integrity-restoration actions are limited to tuple deletions.  We
identify two basic computational issues: {\em repair checking} (is a
database instance a repair of a given database?) and {\em consistent
query answers} \cite{ArBeCh99} (is a tuple an answer to a given query
in every repair of a given database?). We study the computational
complexity of both problems, delineating the boundary between the
tractable and the intractable.  We consider denial constraints,
general functional and inclusion dependencies, as well as key and
foreign key constraints. Our results shed light on the computational
feasibility of minimal-change integrity maintenance. The tractable
cases should lead to practical implementations. The intractability
results highlight the inherent limitations of any integrity
enforcement mechanism, e.g., triggers or referential constraint
actions, as a way of performing minimal-change integrity maintenance.
\end{abstract}
\section{Introduction}
Inconsistency is a common phenomenon in the database world today.
Even though integrity constraints successfully capture data semantics, the
actual data in the database often fails to satisfy such constraints.
This may happen because the data is drawn from a variety
of independent sources (as in data integration \cite{Len02})
or is involved in complex, long-running
activities like workflows.

How to deal with inconsistent data? The traditional way is not to
allow the database to become inconsistent by aborting updates or
transactions leading to integrity violations. We argue that in
present-day applications this scenario is becoming increasingly
impractical. First, if a violation occurs because of data from
multiple, independent sources being merged \cite{LiMe96}, there is no single update
responsible for the violation.  Moreover, the updates have typically
already committed.  For example, if we know that a person should have
a single address but multiple data sources contain different addresses
for the same person, it is not clear how to fix this violation through
aborting some update. Second, the data may have become inconsistent
through the execution of some complex activity and it is no longer
possible to trace the inconsistency to a specific action.

In the context of triggers or referential integrity, more sophisticated
methods for handling integrity violations have been developed. For example,
instead of being aborted an update may be propagated. In general, the
result is at best a consistent database state, typically with no guarantees
on its distance from the original, inconsistent state (the research reported in \cite{LuMaLa97}
is an exception).

In our opinion, integrity restoration should be a separate
process that is executed after an inconsistency is detected. The
restoration should have a minimal impact on the database by trying
to preserve as many tuples as possible. This scenario is
called from now on {\em minimal-change integrity maintenance}.

One can interpret the postulate of minimal change in several different
ways, depending on whether the information in the database is assumed to be {\em correct}
and {\em complete}.  
If the
information is complete but not necessarily correct (it may violate integrity
constraints), the only way to fix the database is by {\em deleting}
some parts of it. If the information is both incorrect and
incomplete, then both insertions and deletions should be considered.  In
this paper we focus on the first case.
Since we are working in the context of the relational data model, we consider
{\em tuple deletions}.
Such a scenario is common in data warehouse
applications where dirty data coming from many sources is cleaned in
order to be used as a part of the warehouse itself.  On the other
hand, in some data integration approaches, e.g.,\cite{Len02,LeLeRo02}, the
completeness assumption is not made.  For large classes of
constraints, e.g., denial constraints, the restriction to deletions
has no impact, since only deletions can remove integrity violations.
We return to the issue of minimal change in Section
\ref{sec:basic}.

We claim that a central notion in the context of integrity restoration
is that of a {\em repair} \cite{ArBeCh99}.  A repair is a database instance that
satisfies integrity constraints and minimally differs from the
original database (which may be inconsistent).  
Because we consider only tuple deletions as ways to restore database
consistency, the repairs in our framework are {\em subsets} of the
original database instance.

The basic computational problem in this context is {\em repair checking}, namely checking
whether a given database instance is a repair of of the original
database. The complexity of this problem is studied in the present
paper.  The PTIME algorithms for repair checking given here can be
easily adapted to non-deterministically compute repairs (as we show).

Sometimes when the data is obtained online from multiple, autonomous
sources, it is not possible to restore the consistency. In that case
one has to settle for computing, in response to queries, {\em
consistent query answers} \cite{ArBeCh99}, namely answers that are
true in every repair of the given database. Such answers constitute a
conservative ``lower bound'' on the information present in the
database. The problem of computing consistent query answers is the
second computational problem studied in the present paper.  We note
that the notion of consistent query answer proposed in \cite{ArBeCh99}
has been used and extended, among others, in
\cite{ArBeCh00,GrGrZu01,LeLeRo02,ABCHRS03,Wij03}.  However, none of
these papers presents a comprehensive and complete computational
complexity picture.

We describe now the setting of our results. 
We analyze the computational complexity of repair checking and consistent query answers
along several different dimensions. We characterize the impact of the following parameters:
\begin{itemize}
\item the {\em class of queries}: quantifier-free queries, conjunctive queries,
and simple conjunctive queries (conjunctive queries without repeated relation symbols).
\item the {\em class of integrity constraints}: 
denial constraints, functional dependencies (FDs), inclusion dependencies (INDs),
and FDs and INDs together. We also consider practically important subclasses of FDs and INDs:
{\em key} functional dependencies and {\em foreign key} constraints \cite{Date81}.
\item the {\em number} of integrity constraints.
\end{itemize}

As a result we obtain several new classes for which both repair checking and consistent query answers
are in PTIME:
\begin{itemize}
\item queries: ground quantifier-free,
constraints: arbitrary denial;
\item queries: closed simple conjunctive, constraints: functional dependencies (at most one FD per relation);
\item queries: ground quantifier-free or closed simple conjunctive, constraints: key functional dependencies 
and foreign key constraints, with at most one key per relation.
\end{itemize}
Additionally, we show that repair checking (but not consistent query answers) 
are in PTIME for arbitrary FDs and acyclic INDs.
The results obtained are tight in the sense that
relaxing any of the above restrictions leads to co-NP-hard problems,
as we prove. (This, of course, does not preclude the possibility that introducing {\em additional}, orthogonal
restrictions could lead to more PTIME cases.)
To complete the picture, we show that
for arbitrary sets of FDs and INDs repair checking is co-NP-complete and 
consistent query answers is $\Pi^p_2$-complete.

Our results shed light on the computational feasibility of minimal-change
integrity maintenance. The tractable cases should lead to practical
implementations. The intractability results highlight the 
inherent limitations of any 
integrity enforcement mechanism, e.g., triggers or referential constraint actions
\cite{MeSi02,LuMaLa97}, as ways of performing minimal-change integrity
maintenance using tuple deletions. 
\ignore{
In our opinion, such results show how surprisingly complex can become in this context
the interactions between relatively simple FDs and INDs.
}

The plan of the paper is as follows.
In Section \ref{sec:basic}, we define the basic concepts.
In Section \ref{sec:denial}, we consider denial constraints.
In Section \ref{sec:ind}, we discuss INDs together with FDs.
In Section \ref{sec:related}, we summarize related research
and in Section \ref{sec:concl} we draw conclusions and discuss future work.
An earlier version of the results in Section \ref{sec:denial} was
presented in \cite{ChMa02a}.

\section{Basic Notions}\label{sec:basic}

In the following we assume we have a fixed relational database schema
$R$ consisting of a finite set of relations. We also have a fixed,
infinite database domain $D$, consisting of uninterpreted constants, and a numeric domain $N$.
Those domains are disjoint.
The database instances can be seen as finite, first-order structures over the given schema, that share the domain $D$.
Every attribute in $U$ is typed, thus all the instances of $R$
can contain only elements either of $D$ or of $N$ in a single attribute.
Since each instance is finite, it
has a finite active domain which is a subset of $D\cup N$.  As usual, we allow
the standard built-in predicates over $N$ ($=,\not=,<,>,\leq,\geq$)
that have infinite, fixed  extensions.
With all these elements we can build
a first order language $\cL$.

\subsection{Integrity Constraints}

Integrity constraints are closed first-order $\cL$-formulas.
In the sequel we will denote relation symbols by $P_1,\ldots,P_m$, 
tuples of variables and constants by $\bar{x}_1,\ldots,\bar{x}_m$, and  
a conjunction of atomic formulas referring to built-in predicates by
$\varphi$.

In this paper we consider the following  basic classes of integrity constraints:
\begin{enumerate}

\item {\em Denial constraints}: $\cL$-sentences 
\[\forall\bar{x}_1,\ldots\bar{x}_k.~ \neg[P_1(\bar{x}_1)\wedge\cdots\wedge P_m(\bar{x}_m)
\wedge \varphi(\bar{x}_1,\ldots,\bar{x}_m)].\]

\item {\em Functional dependencies (FDs)}: $\cL$-sentences
\[\forall \bar{x}_1\bar{x}_2\bar{x}_3\bar{x}_4\bar{x}_5.~ [P(\bar{x}_1,\bar{x}_2,\bar{x}_4) \wedge
P(\bar{x}_1,\bar{x}_3,\bar{x}_5) \Rightarrow \bar{x}_2 = \bar{x}_3],\] where
the $\bar{x}_i$ are sequences of distinct variables. 
A more familiar formulation of the
above FD is $X\rightarrow Y$ where $X$ is the set of attributes of $P$
corresponding to $\bar{x}_1$, and $Y$ the set of attributes of $P$
corresponding to $\bar{x}_2$ (and $\bar{x}_3$).
Clearly, functional dependencies are a special case of denial constraints.

\item {\em Inclusion dependencies (INDs)}: $\cL$-sentences
\[\forall \bar{x}_1~\exists \bar{x}_3.~[Q(\bar{x}_1) \Rightarrow P(\bar{x}_2,
\bar{x}_3)],\] where the $\bar{x}_i$ are sequences of distinct variables with $\bar{x}_2$ contained in $\bar{x}_1$, and $P, Q$
database relations.
Again, this is often written as $Q[Y]\subseteq P[X]$ where $X$ (resp. $Y$)
is the set of attributes of $P$ (resp. $Q$) corresponding to $\bar{x}_2$.
If $P$ and $Q$ are clear from the context, we omit them and write the
dependency simply as $Y\subseteq X$.
{\em Full} inclusion dependencies are those expressible without the existential quantifiers.
\end{enumerate}

Given a set of FDs and INDs $IC$ over a relation $P$
and $X$ which is a key of $P$ w.r.t. $IC$, we say 
that each FD $X\rightarrow Y\in IC$ is a {\em key} dependency
and each IND $Q[Y]\subseteq P[X]\in IC$ is a {\em foreign key constraint}.
If, additionally, $X$ is the primary key of $P$, then both 
kinds of dependencies are termed {\em primary}.

\begin{definition} \label{def:consistency}
Given a database instance $r$ of $R$ and a set of integrity
constraints $\IC$, we say that $r$ is {\em consistent} if $r\vDash
\IC$ in the standard model-theoretic sense; {\em inconsistent}
otherwise. \eop\\
\end{definition}

\subsection{Repairs}
Given a database instance $r$, the {\em set $\Sigma(r)$ of facts} of $r$ is the set of
ground atomic formulas $\{P(\bar{a})~|~r\vDash P(\bar{a})\}$, where $P$ is a relation
name and $\bar{a}$ a ground tuple.

\begin{definition}
The {\em distance} $\Delta^-(r,r')$ between data-base instances $r$ and $r'$ is
defined as
$\Delta^-(r,r') = (\Sigma(r)-\Sigma(r')) .$\eop
\end{definition}

\begin{definition} \label{def:order}
For the instances $r,r',r'' ~$, $r'\leq_r r''$ if
$\Delta^-(r,r')\subseteq\Delta^-(r,r'')$, i.e., if the distance between
$r$ and $r'$ is less than or equal to the distance between $r$ and $r''$.\eop
\end{definition}
\begin{definition}\label{def:repair} 
Given a set of integrity constraints $IC$ and database instances $r$ and $r'$, we say that
$r'$ is a {\em repair} of $r$ w.r.t. $IC$ if $r'\vDash\ IC$ and $r'$ is
$\leq_r$-{\it minimal} in the class of database instances that satisfy $IC$.
\eop
\end{definition}

If $r'$ is a  repair of $r$, then $\Sigma(r')$ is a maximal consistent subset of $\Sigma(r)$.
We denote by $\rpr{IC}{r}$ the set of repairs of $r$ w.r.t. $IC$.
This set is nonempty, since the empty database instance satisfies every set of FDs
and INDs.

\subsection{Queries}

Queries are formulas over the same language $\cL$ as the integrity constraints. 
A query is {\em closed} (or a {\em sentence}) if it has no free variables.
A closed query without quantifiers is also called {\em ground}.
{\em Conjunctive queries} \cite{ChMe77,AbHuVi95} are queries of the form
\[\exists\bar{x}_1,\ldots\bar{x}_m.~[P_1(\bar{x}_1)\wedge\cdots\wedge P_m(\bar{x}_m)
\wedge \varphi(\bar{x}_1,\ldots,\bar{x}_m)].\]
If a conjunctive query has no repeated relation symbols, it is called {\em simple}.

The following definition is standard:
\begin{definition}
A tuple $\bar{t}$ is an {\em answer} to a query
$Q(\bar{x})$ in an instance $r$ iff $r\models Q(\bar{t})$.\eop
\end{definition}

\subsection{Consistent query answers}

Given a query
~$Q(\bar{x})$ to $r$, we want as {\em consistent} answers those tuples that
are unaffected by the violations of ${\it IC}$, even when $r$ violates
${\it IC}$. 

\begin{definition}\label{def:cqa} \cite{ArBeCh99}
A tuple $\bar{t}$ is a {\em consistent answer} to a query
$Q(\bar{x})$ in a database instance $r$ w.r.t. a set of integrity constraints $IC$  iff $\bar{t}$ is an answer to query
$Q(\bar{x})$ in every repair $r^\prime$  of $r$ w.r.t. $IC$. An $\cL$-sentence
$Q$  is {\em consistently true} in $r$ w.r.t. $IC$ if it is true in every repair  of $r$ w.r.t. $IC$. In
symbols:
\[r \models_{IC} Q(\bar{t}) \ \Longleftrightarrow \ 
r^\prime \models Q(\bar{t})\ {\mathit\ for\ every\ repair\ } r^\prime
{\mathit\ of\ } r \ {\mathit w.r.t.}\ IC.\] \eop
\end{definition}

{\em Note:} If the set of integrity constraints $IC$ is clear from the context, we  omit it
for simplicity.

\subsection{Examples}
\begin{example}\label{example:one}
Consider the following instance of a relation {\it Person}
\begin{center}
\begin{tabular}{|lll|}
\hline
{\it Name} & {\it City} & {\it Street}\\\hline
{\rm Brown} & {\rm Amherst} & {\rm 115 Klein}\\
{\rm Brown} & {\rm Amherst} & {\rm 120 Maple}\\
{\rm Green} & {\rm Clarence} & {\rm 4000 Transit}\\\hline
\end{tabular}
\end{center}

\noindent
and the functional dependency ${\it Name}\rightarrow {\it City}\;{\it Street}$.
Clearly, the above instance does not satisfy the dependency.
There are two repairs: one is obtained by removing the first tuple,
the other by removing the second.
The consistent answer to the query $\;{\it Person\/}(n,c,s)\;$
is just the tuple \mbox{\rm (Green,Clarence,4000 Transit)}.
On the other hand, the query $\;\exists s[{\it Person\/}(n,c,s)]\;$
has two consistent answers: \mbox{\rm (Brown,Amherst)} and 
\mbox{\rm (Green,Clarence)}.
Similarly, the query 
\[{\it Person\/}({\rm Brown},{\rm Amherst},{\rm 115\; Klein})
\vee {\it Person\/}({\rm Brown},{\rm Amherst},{\rm 120\; Maple})\]
has {\em true} as the consistent answer.
Notice that for the last two queries the approach based on removing
all inconsistent tuples and evaluating the original query using
the remaining tuples gives different, less informative results.
\end{example}

\begin{example}\label{e:two}
Consider a database with two relations {\it Employee({\it SSN,Name})} and {\it Manager({\it SSN})}.
There are functional dependencies $SSN\rightarrow Name$ and $Name\rightarrow SSN$, and an inclusion
dependency $Manager[SSN]\subseteq Employee[SSN]$.
The relations have the following instances:
\begin{center}
\begin{tabular}{l@{\hspace{20pt}}l}
\begin{minipage}[t]{2in}
\begin{tabular}{|ll|}
\multicolumn{2}{l}{\it Employee}\\\hline
{\it SSN}&{\it Name}\\\hline
123456789 & Smith \\
555555555 & Jones \\
555555555 & Smith \\\hline
\end{tabular}
\end{minipage}
\begin{minipage}[t]{2in}
\begin{tabular}{|l|}
\multicolumn{1}{l}{{\it Manager}}\\\hline
{\it SSN}\\\hline
123456789\\
555555555\\\hline
\end{tabular}
\end{minipage}
\end{tabular}
\end{center}
The instances do not violate the IND but violate both FDs.
If we consider only the FDs, there are two repairs: one obtained
by removing the third tuple from {\it Employee}, and the other by removing the first two tuples
from the same relation.
However, the second repair violates the IND.
This can be fixed by removing the first tuple from {\it Manager}.
So if we consider all the constraints, there are two repairs:
\begin{center}
\begin{tabular}{l@{\hspace{20pt}}l}
\begin{minipage}[t]{2in}
\begin{tabular}{|ll|}
\multicolumn{2}{l}{\it Employee}\\\hline
{\it SSN}&{\it Name}\\\hline
123456789 & Smith \\
555555555 & Jones \\
\hline
\end{tabular}
\end{minipage}
\begin{minipage}[t]{2in}
\begin{tabular}{|l|}
\multicolumn{1}{l}{{\it Manager}}\\\hline
{\it SSN}\\\hline
123456789\\
555555555\\\hline
\end{tabular}
\end{minipage}
\end{tabular}
\end{center}
and
\begin{center}
\begin{tabular}{l@{\hspace{20pt}}l}
\begin{minipage}[t]{2in}
\begin{tabular}{|ll|}
\multicolumn{2}{l}{\it Employee}\\\hline
{\it SSN}&{\it Name}\\\hline
555555555 & Smith \\\hline
\end{tabular}
\end{minipage}
\begin{minipage}[t]{2in}
\begin{tabular}{|l|}
\multicolumn{1}{l}{{\it Manager}}\\\hline
{\it SSN}\\\hline
555555555\\\hline
\end{tabular}
\end{minipage}
\end{tabular}
\end{center}\eop
\end{example}
\begin{example}
We give here some examples of denial constraints.
Consider the relation {\it Emp} with attributes {\it Name}, {\it Salary},
and {\it Manager}, with {\it Name} being the primary key. 
The constraint that {\it no employee can have a salary greater that that of
her manager} is a denial constraint:
\[\forall n,s,m,s',m'.~ \neg [{\it Emp\/}(n,s,m)\wedge {\it Emp\/}(m,s',m')\wedge
s>s'].\]
Similarly, single-tuple constraints ({\tt CHECK} constraints in SQL2) are
a special case of denial constraints. For example, the constraint that
{\em no employee can have a salary over \$200000} is expressed as:
\[\forall n,s,m.~ \neg [{\it Emp\/}(n,s,m)\wedge s>200000].\]
Note that a single-tuple constraint always leads to a single repair which
consists of all the tuples of the original instance that satisfy the constraint.
\end{example}

\subsection{Different notions of repair}

The original notion of repair introduced in \cite{ArBeCh99} required
that the {\em symmetric} difference between a database and its repair be minimized.
As explained in the introduction, this was based on the assumption that the
database may be not only inconsistent but also incomplete.
The notion of repair pursued in the current paper (Definition \ref{def:repair})
reflects the assumption that the database is complete.
There are several reasons for this change of perspective.
First, for denial constraints integrity
violations can only be removed by deleting tuples, so the different notions of repair in fact coincide
in this case.
Therefore, all the results presented in Section \ref{sec:denial} are not affected by the
restriction of the repairs to be subsets of the original instance.
Insertions can restore integrity
only for inclusion dependencies (or, in general for tuple-generating dependencies \cite{AbHuVi95}).
Second, even for inclusion dependencies current language standards like SQL:1999 allow 
only deletions in their repertoire of referential integrity actions.
Third, disallowing insertions significantly strengthens the notion of consistent query answer,
as demonstrated by the following example.
\begin{example}
Consider a database schema consisting of two relations $P(A B)$ and $S(C)$.
The integrity constraints are: the FD $A\rightarrow B$ and the IND $B\subseteq C$.
Assume the database instance $r_1$ consists of $p=\{(a,b),(a,c)\}$ and $s=\{b\}$.
Then under Definition \ref{def:repair} there is only one repair $r_2$ consisting of $p'=\{(a,b)\}$ and $s'=s$.
On the other hand, under the definition of \cite{ArBeCh99}, there is one more repair $r_3$
consisting of $p''=\{(a,c)\}$ and $s''=\{b,c\}$.
Therefore, in the first case $P(a,b)$ is consistently true in the original instance $r_1$,
while in the second case it is not.
Note that $P(a,c)$ is not consistently true in $r_1$ either.
Thus, in the second case $P(a,b)$ and $P(a,c)$ 
are treated symmetrically from the point of view of consistent query answering.
However, intuitively there is a difference between them.
Think of $A$ being the person's name, $B$ her address and $S$ a list of valid addresses.
Then only under Definition \ref{def:repair} would the single valid address be returned as a consistent answer.
\end{example}
Finally, insertions may lead to infinitely many repairs which are, moreover, not very
intuitive as ways of fixing an inconsistent database.
\begin{example}\label{e:two:2}
In Example \ref{e:two}, allowing insertions gives additionally infinitely many repairs 
of the form
\begin{center}
\begin{tabular}{l@{\hspace{20pt}}l}
\begin{minipage}[t]{2in}
\begin{tabular}{|ll|}
\multicolumn{2}{l}{\it Employee}\\\hline
{\it SSN}&{\it Name}\\\hline
123456789 & $c$ \\
555555555 & Smith \\
\hline
\end{tabular}
\end{minipage}
\begin{minipage}[t]{2in}
\begin{tabular}{|l|}
\multicolumn{1}{l}{\it Manager}\\\hline
{\it SSN}\\\hline
123456789\\
555555555\\\hline
\end{tabular}
\end{minipage}
\end{tabular}
\end{center}
where $c$ is an arbitrary string different from {\em Smith}.
\end{example}
\subsection{Computational Problems}

Assume a class of databases ${\cal D}$, a class of queries ${\cal Q}$ and a
class of integrity constraints ${\cal C}$ are given.
We study here the complexity of the following problems:
\begin{itemize}
\item {\em repair checking}, i.e., the complexity of the set 
\[B_{IC}=\{(r,r'):r,r'\in {\cal D} \wedge r'\in\rpr{IC}{r}\},\]
\item {\em consistent query answers}, i.e., the complexity of the set
\[D_{IC,\Phi}=\{r:r\in {\cal D} \wedge r\models_{IC}\Phi\}, \]
\end{itemize}
for a fixed sentence 
$\Phi\in{\cal Q}$ and a fixed finite set $IC\in{\cal C}$ of integrity constraints.
This formulation is called {\em data complexity} \cite{ChHa80,Var82},
since it captures the complexity of a problem as a function of the number 
of tuples in the database instance only. The database schema, the query and
the integrity constraints are assumed to be fixed.

It is easy to see that even under a single key FD, there may be
exponentially many repairs and thus the approach to computing consistent
query answers by generating and examining all repairs is not feasible.

\begin{example}\label{ex:expo}
Consider the functional
dependency $A\rightarrow B$ and
the following  family of relation instances
$r_n$, $n>0$, each of which has $2n$ tuples (represented as columns)
and $2^n$ repairs:
\vspace{-0.6cm}
\begin{center}
\[\begin{array}{c|ccccccc}
r_n & &&&&&&\\ \hline
A & a_1& a_1& a_2& a_2 & ~~\cdots &a_n &a_n\\ \hline
B & b_0& b_1& b_0& b_1 & ~~\cdots &b_0 &b_1 \\ \hline
\end{array}\]
\end{center}
\end{example}

We establish below a general relationship between the problems of repair checking
and consistent query answers. 

\begin{theorem}
In the presence of foreign key constraints, the problem of repair checking is logspace-reducible
to the complement of the problem of consistent query answers.
\end{theorem}
\begin{proof}
We discuss here the case of the database consisting of a single relation $R_0$.
Assume $r$ is the given instance of $R_0$ and 
$r'$ is an another instance of $R_0$ satisfying the set of integrity 
constraints $IC$. We define a new relation $S_0$ having the same attributes as $R_0$
plus an additional attribute $Z$. Consider an instance $s$ of $S_0$ built as follows:
\begin{itemize}
\item for every tuple $(x_1,\ldots,x_k)\in r'$, we add the tuple $(x_1,\ldots,x_k,c_1)$ to $s$;
\item for every tuple $(x_1,\ldots,x_k)\in r-r'$, we add the tuple $(x_1,\ldots,x_k,c_2)$ to $s$.
\end{itemize}
Consider also another relation $P$ having a single attribute $W$ , and a foreign key constraint
$i_0: P[W]\subseteq S_0[Z]$.
The instance $p$ of $P$ consists of a single tuple $c_2$.
We claim that $P(c_2)$ is consistently true in the database instance consisting of $s$ and $p$
w.r.t. $IC\cup\{i_0\}$ iff $r'$ is not a repair of $r$ w.r.t. $IC$.
\end{proof}

\section{Denial constraints}\label{sec:denial}

\subsection{Conflict hypergraph}
Given a set of denial constraints $F$ and an instance $r$, all the repairs of $r$
with respect to $F$ can be succinctly represented as the {\em conflict hypergraph}.
This is a generalization of the {\em conflict graph} defined in \cite{ArBeCh01}
for FDs only.
\begin{definition}\label{d:hypergraph}
The {\em conflict hypergraph} ${\cal G}_{F,r}$ is a hypergraph
whose set of vertices is the set $\Sigma(r)$ of  facts of an instance $r$ and 
whose set of edges consists of all the sets 
\[\{P_1(\bar{t}_1),P_2(\bar{t}_2),\ldots P_l(\bar{t}_l)\}\]
such that $P_1(\bar{t}_1),P_2(\bar{t}_2),\ldots P_l(\bar{t}_l)\in \Sigma(r)$, and there is a constraint
\[\forall \bar{x}_1,\bar{x}_2,\ldots \bar{x}_l.\ 
\neg [P_1(\bar{x}_1) \wedge P_2(\bar{x}_2) \wedge \ldots \wedge P_l(\bar{x}_l)\wedge 
\varphi(\bar{x}_1,\bar{x}_2,\ldots \bar{x}_l)]\]
in $F$ such that $P_1(\bar{t}_1),P_2(\bar{t}_2),\ldots P_l(\bar{t}_l)$ 
violate together  this constraint, which means that there exists 
a substitution $\rho$ such that 
$\rho(\bar{x}_1)=\bar{t}_1, \rho(\bar{x}_2)=\bar{t}_2, \ldots \rho(\bar{x}_l)=\bar{t}_l$ and that 
$\varphi(\bar{t}_1,\bar{t}_2,\ldots \bar{t}_l) $ is true.
\end{definition}

Note that there may be edges in ${\cal G}_{F,r}$ that contain only one vertex. 
Also, the size of the conflict hypergraph is polynomial in the number of tuples in the 
database instance.

By an {\em independent set} in a hypergraph we mean a subset of its set of vertices
which does not contain any edge.  

\begin{proposition}\label{p:indset}
Each repair of $r$ w.r.t. $F$  corresponds to
a maximal independent set in ${\cal G}_{F,r}$.
\end{proposition}

Proposition \ref{p:indset} yields the following result:

\begin{proposition}\cite{ABCHRS03}\label{p:fds}
For every set of denial constraints $F$ and $\cL$-sentence $\Phi$,
$B_F$ is in PTIME and $D_{F,\Phi}$ is in co-NP.\eop
\end{proposition}

Note that the repairs of an instance $r$ can be computed nondeterministically
by picking a vertex of ${\cal G}_{F,r}$ which does not belong to a single-vertex edge 
and adding vertices that do not result in the addition of an entire edge.

\subsection{Positive results}
A set of constraints is {\em generic} if it does not imply any ground literal.
The results in \cite{ArBeCh99} imply the following:
\begin{proposition}\label{prop:pods99}
For every generic set $F$ of binary denial constraints  and full inclusion dependencies, 
and quantifier-free $\cL$-sentence 
\[\Phi=P_1(\bar{x}_1)\wedge\cdots P_m(\bar{x}_m)\wedge \neg P_{m+1}(\bar{x}_{m+1})\wedge\cdots \wedge\neg P_n(\bar{x}_n)\wedge
\varphi(\bar{x}_1,\ldots,\bar{x}_n),\]
$D_{F,\Phi}$ is in PTIME.\eop
\end{proposition}

The techniques in \cite{ArBeCh99} do not generalize to non-binary constraints, or
queries involving disjunction or quantifiers. However, non-binary constraints and disjunctions do not necessarily lead to
intractability, as shown by the following theorem.

\begin{theorem}\label{th:qf}
For every set $F$  of denial constraints  and quantifier-free $\cL$-sentence $\Phi$, $D_{F,\Phi}$ is in PTIME.\eop
\end{theorem}
\begin{proof}
We assume the sentence is in CNF, i.e., of the form
$\Phi = \Phi_1\wedge \Phi_2 \wedge \ldots \Phi_l $, where  each $\Phi_i $ is a disjunction of
ground literals. $\Phi$ is true in every repair of $r$ if and only if each of the clauses $\Phi_i $ is
true in every repair. So it is enough to provide a polynomial algorithm which will check if   
a given  ground clause is consistently {\em true}.

It is easier to think that we are checking if a ground clause {\em true} is {\bf not\/} consistently true.
This means that we are checking, whether there exists a repair $r'$ in which $\neg \Phi_i$ is true
for some $i$. 
But $\neg \Phi_i $ is of the form 
$ P_1(\bar{t}_1)\wedge P_2(\bar{t}_2)  \wedge \ldots \wedge  
P_m(\bar{t}_m)\wedge \neg P_{m+1}(\bar{t}_{m+1})\wedge\ldots\wedge \neg 
P_n(\bar{t}_n)$, 
where the $\bar{t}_j$'s are  tuples of constants.
WLOG, we assume that all the facts in the set $\{P_1(\bar{t}_1),\ldots,P_n(\bar{t}_n))\}$
are mutually distinct.

The nonderministic algorithm selects for every $j$, $m+1\leq j \leq n$, $\bar{t}_j\in r$,
an edge $E_j\in {\cal G}_{F,r}$ such that $\bar{t}_j\in E_j$.
Additionally the following global condition needs to be satisfied:
there is no edge $E\in {\cal G}_{F,r}$ such that
$E\subseteq {r'}$
where
\[{r'}= \{\bar{t}_1,\ldots,\bar{t}_m\}\cup \bigcup_{m+1\leq j\leq n,\bar{t}_j\in r} (E_j-\{\bar{t}_j\}).\]

If the selection succeeds, then a repair in which $\neg \Phi_i$ is  true can be built by adding to $r'$ new tuples 
from $r$ until the set is maximal independent. 
The algorithm  needs  $n-m$ nondeterministic steps, a number which is
independent of
the  size of the database (but dependent on $\Phi$), 
and in each of its nondeterministic steps selects one possibility from a
set
whose size is polynomial in the size of the database. 
So there is an equivalent PTIME deterministic algorithm.
\end{proof} 
 
In the case when the set $F$ of integrity constraints consists of only one FD per relation the
conflict hypergraph has a very simple form.  It is a disjoint union of full multipartite graphs. 
If this single dependency is a key dependency then the conflict graph is a union of disjoint cliques. 
Because of this very simple structure we hoped that it would be possible, in such a situation, 
to compute in polynomial time the consistent answers not only to
ground queries, but also to all  conjunctive queries. As we are going to see now, this is only possibly
if the conjunctive queries are suitably restricted.

\begin{theorem}\label{th:conj}
Let $F$ be a set of FDs, each dependency over a different relation
among $P_1,P_2,\ldots,P_k$.
Then for each closed simple conjunctive query  $Q$, 
\ignore{
that refers to the relations $P_1,P_2,\ldots,P_k$ only,
  of the form 
$\exists \bar{x}_1,\bar{x}_2,\ldots,\bar{x}_k 
[P_1(\bar{x}_1)\wedge P_2(\bar{x}_2)\wedge\ldots\wedge P_k(\bar{s}_k)\wedge \varphi(\bar{x}_1,\bar{x}_2,\ldots \bar{x}_l)]$
(where  $\varphi$ is quantifier-free and only built-in predicates occur there and all the variables in the first part of the
formula are distinct),
}
there exists a sentence $Q'$ such that for every database instance {r},
$r\models_F Q$ 
iff $r\models Q'$.
Consequently, $D_{F,Q}$ is 
in PTIME.
\end{theorem}

\begin{proof}
We present the construction for $k=2$ for simplicity;
the generalization to an arbitrary $k$ is straightforward.
Let $P_1$ and $P_2$ be two different relations of arity $k_1$ and $k_2$, resp.
Assume we have the following FDs: $Y_1\rightarrow Z_1$ over $P_1$ and
$Y_2\rightarrow Z_2$ over $P_2$.
Let $\bar{y}_1$ be a vector of arity $|Y_1|$, $\bar{y}_2$ a vector of arity $|Y_2|$,
$\bar{z}_1$ and $\bar{z}_1'$ vectors of arity $|Z_1|$, and $\bar{z}_2$ and $\bar{z}_2'$ vectors of arity $|Z_2|$.
Finally, let $\bar{w}_1,\bar{w}_1',\bar{w}_1''$ (resp. $\bar{w}_2,\bar{w}_2',\bar{w}_2''$)
be vectors of arity $k_1-|Y_1| - |Z_1|$ (resp. $k_2-|Y_2| - |Z_2|$).
All of the above vectors consist of distinct variables.
The query $Q$ is of the following form
\[\exists\bar{y}_1,\bar{z}_1,\bar{w}_1,\bar{y}_2,\bar{z}_2,\bar{w}_2.~
[P_1(\bar{y}_1,\bar{z}_1,\bar{w}_1)\wedge P_2(\bar{y}_2,\bar{z}_2,\bar{w}_2)
\wedge\varphi(\bar{y}_1,\bar{z}_1,\bar{w}_1,\bar{y}_2,\bar{z}_2,\bar{w}_2)].\]
Then, the query $Q'$ is as follows:
\[\begin{array}{l}
\exists\bar{y}_1,\bar{z}_1,\bar{w}_1,\bar{y}_2,\bar{z}_2,\bar{w}_2
\forall \bar{z}_1',\bar{w}_1',\bar{z}_2',\bar{w}_2'\exists\bar{w}_1'',\bar{w}_2''
[P_1(\bar{y}_1,\bar{z}_1,\bar{w}_1)\wedge P_2(\bar{y}_2,\bar{z}_2,\bar{w}_2)
\wedge\varphi(\bar{y}_1,\bar{z}_1,\bar{w}_1,\bar{y}_2,\bar{z}_2,\bar{w}_2)\\
\wedge (P_1(\bar{y}_1,\bar{z}_1',\bar{w}_1')\wedge P_2(\bar{y}_2,\bar{z}_2',\bar{w}_2')
\Rightarrow P_1(\bar{y}_1,\bar{z}_1',\bar{w}_1'')\wedge P_2(\bar{y}_2,\bar{z}_2',\bar{w}_2'')
\wedge\varphi(\bar{y}_1,\bar{z}_1',\bar{w}_1'',\bar{y}_2,\bar{z}_2',\bar{w}_2''))].
\end{array}\]
\end{proof}

We show now that the above results are the strongest possible, since
relaxing any of the restrictions leads to co-NP-completeness.
This is the case even though we limit ourselves to {\em key} FDs.

\subsection{One key dependency, nonsimple conjunctive query}

\ignore{
\begin{lemma}\label{zksiazeczki}
3SAT is NP-complete even if restricted only to instances, in which every clause
either contains only positive literals or only negative literals.
\end{lemma}
\begin{proof} For each variable $p$ 
substitute every occurrence of $\neg p$ by a new variable $p'$. Then add clauses
$p\vee p'$ and $\neg p\vee \neg p'$.\end{proof}
}

\begin{theorem}\label{bliskosiebie}
There exist a key FD $f$ and a closed conjunctive query 
\[Q\equiv\exists x,y,z.~ [R(x,y,c)\wedge R(z,y,c')],\]
for which $D_{\{f\},Q}$ is co-NP-complete.
\end{theorem}

\begin{proof}
Reduction from MONOTONE 3-SAT.
The FD is $A\rightarrow BC$.
Let $\Phi=\phi_1\wedge\ldots \phi_m \wedge \psi_{m+1} \ldots \wedge\psi_l $ be a conjunction of clauses, such that
all occurrences of variables in $\phi_i$ are positive and all occurrences of variables in $\psi_i$ are negative.
We build a database with  the facts $R(i,p,c)$ if the variable $p$ occurs in the clause $\psi_i$
and  $R(i,p,c')$ if the variable $p$ occurs in the clause $\phi_i$. 
Now, there is an assignment which satisfies $\Phi$  
if and only if  there exists a repair 
of the database in which $Q$ is false. 
To show  the $\Rightarrow$ implication, select for each clause $\phi_i$ one variable $p_i$
which occurs in this clause and whose value is 1 and for each clause $\psi_i$  one variable $p_i$
which occurs in $\psi_i$ and whose value is 0. The set of facts 
$\{R(i,p_i,c):i\leq m\}\cup  \{R(i,p_i,c'):m+1\leq i\leq l\}$ is a repair in which the query $Q$ is false. 
The $\Leftarrow$ implication is even simpler.
\end{proof}

\subsection{Two key dependencies, single-atom query}

By a {\em bipartite edge-colored graph} we mean  a tuple ${\cal G}=\langle V,E,B,G\rangle $ such that
 $\langle V,E\rangle$ is an undirected bipartite graph and  $E=B\cup G $ for some given 
 disjoint  sets $B,G$ (so we think that each of the edges of $\cal G$ has one
of the two colors). 

\begin{definition}
Let  ${\cal G}=\langle V,E,B,G\rangle $ be a  bipartite edge-colored graph, 
and let $M\subset E$.  
We say that $M$ is maximal $\cal V$-free if:

\begin{enumerate}
     \item $M$ is a maximal (w.r.t. inclusion) subset of $E$ with the property
      that neither $M(x,y)\wedge M(x,z)$ nor $M(x,y)\wedge M(z,y)$ holds
      for any $x,y,z$.

     \item $M\cap B =\emptyset$.
\end{enumerate} 

We say that $\cal G$ 
has the max-$\cal V$-free property
if there exists $M$ which is maximal $\cal V$-free.
\end{definition}

\begin{lemma}\label{2key}
Max-$\cal V$-free is an NP-complete property of bipartite edge-colored graphs.
\end{lemma}
\begin{proof}
Reduction from 3-COLORABILITY. Let ${\cal H}=\langle U,D\rangle $ be some 
undirected graph. 
This is how we define the bipartite edge-colored graph  ${\cal G}_{\cal H}$:

\begin{enumerate}
\item
$V=\{v_\varepsilon, v'_\varepsilon :v\in U,\varepsilon\in\{m,n,r,g,b\}\}$,
which means that there are 10 nodes in the graph $\cal G$ for each node of $\cal H$;

\item
$G(v_m,v'_r),G(v_m,v'_b),G(v_n,v'_b),G(v_n,v'_g)$ and
$G(v_r,v'_m)$,$G(v_b,v'_m)$,$G(v_b,v'_n)$,$G(v_g,v'_n)$ hold for each $v\in U$;

\item
$B(v_\epsilon,v'_\varepsilon) $ holds for each $v\in U$ and each pair 
$\epsilon,\varepsilon\in\{r,g,b\}$ such that $\epsilon \neq \varepsilon $;

\item
$B(v_\varepsilon,u'_\varepsilon) $ holds for each $\varepsilon\in \{r,g,b\}$
and each  pair $u,v\in U$  such that $ D(u,v) $.

\end{enumerate}

Suppose that $\cal H$ is 3-colorable. We fix a coloring of $\cal H$ and construct the set $M$.
 For each $v\in U$: if the color of $v$ is Red, then  the edges 
$G(v_m,v'_b),G(v_n,v'_g)$ and $G(v_b,v'_m),G(v_g,v'_n)$ are in $M$.
If color of $v$ is Green, then  
 the edges 
$G(v_m,v'_r),G(v_n,v'_b)$ and $G(v_r,v'_m), G(v_b,v'_n)$
are in $M$, and if  the color of $v$ is Blue, then  
 the edges 
$G(v_m,v'_r),G(v_n,v'_g)$ and $G(v_r,v'_m),G(v_g,v'_n)$ are in $M$. It is easy to see that 
the set $M$ constructed in this way is maximal $\cal V$-free.

For the other direction, suppose that a maximal $\cal V$-free set $M$ exists in ${\cal G}_{\cal H}$.
Then, for each $v\in U$ there is at least one  node among $v_r,v_g, v_b$ which does not belong 
to any $G$-edge in $M$. Let $v_\epsilon$ be this node.
Also, there is at least one such node (say, $v'_\varepsilon$) among $v'_r,v'_g, v'_b$. Now, it follows easily
from the construction of ${\cal G}_{\cal H}$ that if $M$ is  
maximal $\cal V$-free then $\epsilon = \varepsilon$. Let this $\epsilon $ be color of $v$ in $\cal G$. 
It is easy to check that the coloring  defined in this way is a legal 3-coloring of $\cal G$.
\end{proof}

\begin{theorem}
There is a set $F$ of two key dependencies and a closed conjunctive query $Q\equiv\exists x,y.~ [R(x,y,b)]$, 
for which $D_{F,Q}$ is co-NP-complete.
\end{theorem}

\begin{proof}
The 2 dependencies are $A\rightarrow BC$ and $B\rightarrow AC$. For a given 
 bipartite edge-colored graph  ${\cal G}=\langle V,E,B,G\rangle $ we build a database with 
 the tuples $(x,y,g)$ if $G(x,y) $ holds in ${\cal G}$ and $(x,y,b)$ if $B(x,y) $ holds in ${\cal G}$.
 Now the theorem follows from Lemma \ref{2key} since
 a repair in which  the query $Q$ is not true exists if and only if 
 ${\cal G}$ has the max-$\cal V$-free property. \end{proof}

\subsection{One denial constraint}

By an {\em edge-colored graph} we mean  a tuple ${\cal G}=\langle V,E,P,G,B\rangle $ such that
 $\langle V,E\rangle$ is a (directed) graph and  $E=P\cup G \cup B$ for some given 
pairwise disjoint  sets $P,G,B$ (which we interpret as colors). 
We say that the edge colored graph $\cal G$ has the $\cal Y$ property if there are 
$x,y,z,t\in E$ such that 
$E(x,y),E(y,z),E(y,t)$ hold and the edges $E(y,z)$ and $E(y,t)$ are of different colors.

\begin{definition}\label{tenF}
We say that the edge-colored graph $\langle V,E,P,G,B\rangle$ has the max-$\cal Y$-free property
if there exists a subset $M$ of $E$ such that $M\cap P =\emptyset$ and :
\begin{enumerate}
\item $\langle V,M,P\cap M,G\cap M,B\cap M \rangle $ does not have the $\cal Y$-property;

\item $M$ is a maximal (w.r.t. inclusion) subset of $E$ satisfying the first condition;

\end{enumerate} 
\end{definition}

\begin{lemma}\label{1denial}
Max-$\cal Y$-free is an NP-complete property of edge-colored graphs.
\end{lemma}
\begin{proof}
By a reduction of 3SAT. Let $\Phi= \phi_1\wedge \phi_2\wedge \ldots \wedge \phi_l$ be 
conjunction of clauses. Let $p_1,p_2,\ldots p_n$ be all the variables in $\Phi$.
This is how we define the  {\em edge-colored graph} ${\cal G}_{\Phi}$: 

\begin{enumerate}
\item 
$V=\{a_i,b_i,c_i,d_i:1\leq i\leq n\}\cup \{e_i,f_i,g_i:1\leq i\leq l\}$,
which means that there are 3 nodes in the new graph for each clause in $\Phi$
and 4 nodes for each variable. 

\item $P(a_i,b_i)$ and $P(e_j,f_j)$ hold for each suitable $i,j$;

\item $G(b_i,d_i)$ and $G(e_j,g_j)$ hold for each suitable $i,j$;

\item $B(b_i,c_i)$  holds for each suitable $i$;

\item $G(d_i,e_j)$ holds if $p_i$ occurs positively in $\phi_j$;

\item $B(d_i,e_j)$ holds if $p_i$ occurs negatively in $\phi_j$;

\item  $E=B\cup G\cup P$.
\end{enumerate}

Now suppose that $\Phi $ is satisfiable, and that $\mu$ is the 
satisfying assignment. We define the set $M\subset E$ as follows.
We keep in $M$ all the $G$-colored edges from item 3 above. If $\mu(p_i)=1$
then we keep in $M$ all the $G$ edges leaving $d_i$ (item 5). Otherwise we  
keep in $M$ all the $B$ edges leaving $d_i$ (item 6). Obviously, $M\cap P=\emptyset$.
It is also easy to see that $M$ does not have the $\cal Y$-property and that it is maximal.

In the opposite direction, notice that if an $M$, as in Definition \ref{tenF} does exist, then
it  must contain
all the $G$-edges from item 2 above - otherwise a $P$ edge could be added without 
leading to the $\cal Y$-property. But this means that, for each $i$,
$M$ can either contain some (or all) of the $B$-edges leaving $d_i$ or some (or all) of the $G$-edges.
In this sense $M$ defines a valuation of variables. Also, if $M$ is maximal, it must contain,
for each $j$, at least one edge leading to $e_j$. But this  means that the defined valuation 
satisfies $\Phi$.
\end{proof}


\begin{theorem}
There exist a denial constraint $f$ and a closed conjunctive query 
\[Q\equiv\exists x,y.~ [R(x,y,p)],\] 
for which $D_{\{f\},Q}$ is co-NP-complete.
\end{theorem}

\begin{proof}
The denial constraint $f$ is:
\begin{center}
 $\forall x,y,z,s,s',s''\; \neg[ R(x,y,s) \wedge R(y,z.s') \wedge R(y,w,s'')\wedge s'\neq s'']$
\end{center}
 For a given  edge-colored graph  ${\cal G}=\langle V,E,P,G,B\rangle $ we build a database with 
 the tuples $R(x,y,g)$ if $G(x,y) $ holds in ${\cal G}$, with  $R(x,y,p)$ if $P(x,y) $ holds in ${\cal G}$
 and with $R(x,y,b)$ if $B(x,y) $ holds in ${\cal G}$.
 Now the theorem follows from Lemma \ref{1denial} since
 a repair in which  the query $Q$ is not true exists iff
 ${\cal G}$ has the max-$\cal Y$-free property. \end{proof}

\section{Inclusion dependencies}\label{sec:ind}
\begin{proposition}\label{p:inds}
For every set of INDs $I$ and $\cL$-sentence $\Phi$,
$B_I$ and $D_{I,\Phi}$ are in PTIME.
\end{proposition}
\begin{proof}
For a given database instance $r$, a single repair is obtained by deleting all the tuples
violating $I$ (and only those).
\end{proof}

We consider now FDs and INDs together.

\subsection{Single-key relations}\label{sec:single}

We want to identify here the cases where both repair checking and computing consistent query answers
can be done in PTIME. The intuition is to limit the interaction between the FDs and the INDs in the 
given set of integrity constraints in such a way that one can use the PTIME results obtained for
FDs in the previous section and in \cite{ABCHRS03}. 

\begin{lemma}\label{l:onekey}
Let $IC=F\cup I$ be a set of constraints consisting of a set of
key FDs $F$ and  a set of foreign key constraints $I$ but with no more than one
key per relation.  Let $r$ be a database instance and $r'$ be the unique 
repair of $r$ with respect to the foreign key constraints  in $I$. Then $r''$
is a repair of $r$ w.r.t. $IC$ if and only if it is a repair of $r'$
w.r.t. $F$.
 \end{lemma}
\begin{proof}
The only thing to be noticed here is that  repairing  $r'$ with respect to key 
constraints does not lead to new inclusion violations. This is because the set of key values in each relation remains
unchanged after such a repair (which is not necessarily the case if we have relations with more than one key).
\end{proof}
 
\begin{corollary}
Under the assumptions of Lemma \ref{l:onekey}, $B_{IC}$ is in PTIME.
\end{corollary}
\begin{proof}
Follows from Proposition \ref{p:fds}.
\end{proof}

The repairs w.r.t. $IC=F\cup I$ of $r$ are computed by (deterministically)
repairing $r$ w.r.t. $I$ and then nondeterministically
repairing the result w.r.t. $F$ (as described in the previous section).

We can also transfer the PTIME results about consistent query answers obtained for FDs only.
\begin{corollary}
Let $\Phi$ a quantifier-free $\cL$-sentence or a simple conjunctive closed $\cL$-query.
Then under the assumptions of Lemma \ref{l:onekey}, $D_{IC,\Phi}$ is in PTIME.
\end{corollary}
\begin{proof}
From Theorem \ref{th:qf} and Theorem \ref{th:conj}.
\end{proof}
 
Unfortunately, the cases identified above are the only ones we know of in which both
repair checking and consistent query answers are in PTIME.

\subsection{Acyclic inclusion dependencies}\label{sec:acyclic}

For acyclic INDs (and arbitrary FDs), the repair checking problem is still in PTIME.
Surprisingly, consistent query answers becomes in this case a co-NP-hard problem, even
in the case of key FDs and primary key foreign key constraints. 
If we  relax  any of the assumptions of Lemma \ref{l:onekey}, 
the problem of consistent query answers becomes intractable, even under acyclicity.

\begin{definition}\cite{AbHuVi95}
Let $I$ be a set of INDs over a database schema $R$.
Consider a directed graph whose
vertices are relations from $R$ and such that there is an edge $E(P,R)$ in
the graph if and only if  there is an IND of the form $P[X]\subseteq R[Y]$ in $I$.
A set of inclusion dependencies is {\em acyclic} if the above graph does not have
a cycle.\eop
\end{definition}

\begin{theorem}\label{t:acyclic}
Let $IC=F\cup I$ be a set of constraints consisting of a set of
FDs $F$ and  an acyclic set of INDs $I$.
Then $B_{IC}$ is in PTIME.
\end{theorem}
\begin{proof}
First compare $ r$ and $ r'$ on relations which are not on the left-hand side of any
IND in $I$. Here, $ r'$ is a repair if and only if the functional dependencies 
are satisfied in $ r'$ and if adding to it any additional tuple from $ r$ would violate one of the functional dependencies.
Then consider relations which are on the left-hand side of some INDs, but the inclusions only lead to already checked
relations. Again,  $ r'$ is  a repair of those relations if and only if adding any new tuple (i.e. any tuple from 
$ r$ but not from $ r'$) would violate some constraints. Repeat the last step until all the relations are checked.
\end{proof}

The above proof yields a nondeterministic PTIME procedure for computing the repairs w.r.t. $IC=F\cup I$.

To our surprise, Theorem \ref{t:acyclic} is the strongest possible positive result. The problem of consistent 
query answers is already intractable, even under additional restrictions on the FDs and INDs.
To see this let us start by establishing NP-completeness of the {\em maximal spoiled-free} problem.

By an instance of the maximal spoiled-free problem we will mean 
 ${\cal G}= \langle V,V_1,V_2,V_3,S,E\rangle$  such that:
 
 \begin{enumerate}
 \item  $\langle V,E\rangle $ is a ternary
undirected  hypergraph (so $V$ is a set of vertices and $E$ is a set of {\em triangles});
 
 \item
 $V_1,V_2,V_3$ are pairwise disjoint;
 
 \item
 $V_1 \cup V_2 \cup V_3 = V$;
 
 \item
 Relation $E$ is typed: if $E(a,b,c)$ holds in $\cal G$ then $a\in V_1$, $b\in V_2$ and $c\in V_3$;
 
 \item
 $S\subseteq V_1$ ($S$ will be called set of {\em  spoiled vertices}).
 
 \end{enumerate}

We will consider maximal (with respect to inclusion)
sets of disjoint triangles in $\cal G$.
 We  call a triangle {\em spoiled}
if one of its vertices is spoiled. 
The {\em maximal  spoiled-free} problem is defined  as the problem of deciding, for a 
given  instance ${\cal G}=\langle V,V_1,V_2,V_3,S,E\rangle $,
 if there exists a maximal set $T\subset E$ of disjoint 
triangles, such that none of the triangles in $T$ is spoiled. It is easy to get confused here, so let 
us explain that the problem we are considering here is  not the existence  of a set of disjoint triangles, 
which would be maximal
in the class of sets not containing a spoiled triangle: such a set of course always exists. 
The problem we consider is 
the existence  of a set of disjoint triangles in $\cal G$ which is not only maximal but also 
does not contain a spoiled triangle.

\begin{lemma}
The maximal spoiled-free problem is NP-complete.
\end{lemma}
\begin{proof}
By a reduction of 3-colorability. Let ${\cal H}=\langle U,D\rangle $ be some 
undirected graph. 
We are going to construct the instance of the maximal spoiled-free problem  ${\cal G}_{\cal H}$.
The construction is a little bit complicated, and we hope to simplify the presentation
by the following convention:

Each vertex in $V_1$ belongs to exactly one triangle in $E$. So a triangle is fully specified
by its vertex in $V_2$, its vertex in $V_3$ and by the information if it is spoiled or not.
 
Now, for each vertex $v$ in $U$ we will have vertices 
$v_r,v_g,v_b,v_p,v_q$ in $V_2$ and vertices $v'_r,v'_g,v'_b,v'_p,v'_q$ in $V_3$. 
The only nonspoiled triangles will be the defined by the following pairs:
$[v_r,v'_p]$, $[v_g,v'_p]$, $[v_g,v'_q]$, $[v_b,v'_q]$,
$[v_p,v'_r]$, $[v_p,v'_g]$, $[v_q,v'_g]$, $[v_q,v'_b]$  (so we have 8
nonspoiled triangles for each vertex in $U$).

There are two kinds of spoiled triangles. 
For each $v\in U$, and for each pair $\epsilon,\varepsilon\in \{r,g,b\} $ such that
$\epsilon\neq \varepsilon$  there is a spoiled triangle $[v_\epsilon,v'_\varepsilon]$ in $\cal G$.
For each $v,u\in U$, such that $D(v,u)$ holds in $\cal H$, 
and for each $\epsilon \in \{r,g,b\} $   there is a spoiled triangle 
$[v_\epsilon,u_\epsilon]$ in $\cal G$.

Now we need to show that $\cal H $ is 3-colorable if and only if there 
exists a maximal set $T\subset E$ of disjoint 
triangles, such that none of the triangles in $T$ is spoiled. 

Let us start from the  $\Rightarrow $ direction, which  is simple. Consider a coloring of 
$\cal H$ with colors $r,g$ and $b$. Now take $T$ as a set containing,
for each vertex $v$  of $\cal H$ with some color 
$\epsilon$,  all  nonspoiled triangles of the form $[v_\alpha,v'_\beta]$ where neither $\alpha$ nor
$\beta$ equals to $\epsilon$. 
Obviously, $T$ defined in this way, does not contain spoiled triangles.
A simple analysis shows that it is also maximal.

For the other direction suppose that there is a set $T$ of disjoint triangles in $\cal G$ which is
maximal and only contains nonspoiled triangles. It is easy to see that for each $v$ 
exactly one of the vertices $v_r,v_g,v_b $ is not in any triangle in $T$, and that also
among $v'_r,v'_g,v'_b$ there is exactly one which is not in any triangle in $T$. If 
they were different, in the sense that first of them were $v_\varepsilon$ and the second
$v'_\epsilon$, for $\epsilon\neq \varepsilon$, then  a spoiled triangle $[v_\varepsilon,v'_\epsilon]$
could be added to $T$ what contradicts its maximality. So they are equal, and in a natural way they define a color 
of $v$. Now we need to prove that the coloring of $\cal H$ defined in this way is a legal one. 
But if $D(u,v)$ holds in $\cal H$ then there is spoiled triangle $[v_\epsilon,u_\epsilon]$ in $\cal G$ for
each $\epsilon \in\{r,g,b\}$. So if the colors of $v$ and $u$ were both equal to some $\epsilon$,
 then we could add this spoiled triangle, and $T$ would not be maximal. 
\end{proof}

\begin{theorem}\label{t:acyclic:cqa}
There exist a database schema, a set $IC$ of integrity constraints consisting of key FDs
and of an acyclic set of primary foreign key constraints, and a ground atomic query $\Phi$
such that $D_{IC,\Phi}$ is co-NP-hard.
\end{theorem}
\begin{proof}
The schema consists of a unary relation $P$, a binary relation $Q(Q_1,Q_2)$ 
 and of a ternary relation $R(R_1,R_2,R_3)$. The columns $Q_1$,$R_1$,$R_2$,$R_3$ 
 are keys, with $Q_1$ and $R_1$ being the primary keys.  The foreign key dependencies are $P\subseteq Q_1$
 and $Q_2\subseteq R_1$. 
 For a given instance  ${\cal G}$  of the maximal spoiled-free problem we will 
 construct
 a database instance $r$, and a query $\Phi$ such that  ${\cal G}$ has the 
 maximal spoiled-free property if and only if there is a repair $ r'$ of $ r$ with respect to 
 $IC$ such that $\Phi $ is not true in $ r'$. 
 
 We define the relation $P$ as a single fact $P(a)$. The relation $Q$ is defined as 
 a set of facts $\{Q(a,s): s\in S\}$, where $S$ is the set of spoiled vertices from $\cal G$.
 Finally, $R$ is the hypergraph from $\cal G$. The query $\Phi$ is $P(a)$.
 
 The repairs of $R$ with respect to the key dependencies correspond to maximal sets of 
 disjoint triangles in $\cal G$. 
 If $\cal G$ has the  maximal spoiled-free property then there exists a repair of $R$ which does not
 contain any tuple of the form  $ R(s,u,v) $ with $s\in S$. But then the only way to repair
 $Q$ is it take the empty relation, and, consequently, the only way to repair $P$ is to take
 the empty relation. So if $\cal G$ has the  maximal spoiled-free property then $\Phi $ indeed is not
 true in all repairs. For the other direction notice that if if each repair of $R$ it is a tuple 
 of the form  $ R(s,u,v) $ with $s\in S$ then each repair of $Q$ is nonempty and in consequence
 each repair of $P$ consists of the single atom $P(a)$, so then $\Phi $ is indeed 
 true in all repairs.
\end{proof}

\subsection{Relaxing acyclicity}\label{sec:rest}

We show here that relaxing the acyclicity assumption in Theorem \ref{t:acyclic} leads to the
intractability of the repair checking problem (and thus also the problem of consistent query answers),
even though alternative restrictions on the integrity constraints are imposed.

\subsubsection{One FD, one IND}

\begin{theorem}
There exist a database schema and a set $IC$ of integrity constraints, consisting of one FD
and one IND, such that $B_{IC}$ is co-NP-hard.
\end{theorem}
\begin{proof}
We will check here whether the empty set is a repair.
The database schema consists of one relation $R(A_1,A_2,A_3,A_4)$ and the constraints in $IC$ are 
$A_1\rightarrow A_2$ and $A_3\subseteq A_4$.

Consider a propositional formula $\Phi=\phi_1\wedge \phi_2 \wedge \ldots \phi_m$, where $\phi_i $ are clauses. 
Let $ r_{\Phi}$ consist of 
the facts $R(p_j,0,\phi_{i},\phi_{i+1})$ such that $p_j$ occurs negatively in $\phi_i$ and 
 of 
the facts $R(p_j,1,\phi_{i},\phi_{i+1})$ such that $p_j$ occurs positively in $\phi_i$
where the addition $i+1$ is meant modulo the number $m$ of clauses in $\Phi$.
We want to  show   that $\emptyset$ is a repair of
$ r_{\Phi}$ with respect to $IC$ if and only if $\Phi $ is not satisfiable. 

For the {\em only if } direction notice that if $\rho $ is a satisfying assignment of $\Phi$ then
the subset of $ r_\Phi$ consisting  of all the facts of the form  $R(p,\rho(p),\phi_{i},\phi_{i+1})$ is a repair, 
and obviously $\emptyset $ is not a repair then.

 For the opposite direction first notice that 
a repair $r'$ of $ r_\Phi$ which is nonempty contains some fact of the form $R(\_,\_,\phi_i,\phi_{i+1})$. So, by
inclusion $A_3\subseteq A_4$ it must also contain some fact of the form $R(\_,\_,\phi_{i-1},\phi_{i})$. By induction
we show that
\begin{quote}
(*) for every clause $\phi_j$ from $\Phi$ there is a fact of the form $R(\_,\_,\phi_{j},\phi_{j+1})$
in $ r'$.
\end{quote}
Now we make use of the functional dependency $A_1\rightarrow A_2$. If $ r'$ is a repair of $ r_\Phi$
then for each variable $p$ there are either only facts of the form  $R(p,0,\_,\_)$ in $ r'$ or
only facts of the form $R(p,1,\_,\_)$. Define the assignment $\rho(p)$ as 1 if there is some fact of the form 
$R(p,1,\_,\_)$ in $ r'$  and  as 0 otherwise. It follows from the construction of $ r_\Phi$ that
if a clause of the form $R(\_,\_,\phi_{j},\phi_{j+1})$ is in $ r'$ then $\rho $ satisfies $\phi_j$. Together with (*)
this completes the proof.
\end{proof}

\subsubsection{Key FDs and foreign key constraints}

\begin{theorem}
There exist a database schema and a set $IC$ of integrity constraints, consisting of 
key FDs and  foreign key constraints,  such that $B_{IC}$ is co-NP-hard.
\end{theorem}

\begin{proof}
Again we consider checking whether the empty set is a repair.
The schema consists of 10 binary relations: $R(A,B),R_{i,j}(A_{i,j},B_{i,j})$ with $1\leq i,j\leq 3$.
 For each pair $(i,j)$ both the key dependencies 
 $A_{i,j}\rightarrow B_{i,j}$ and  $ B_{i,j}\rightarrow A_{i,j}$ are in $IC$, with $A_{i,j}$
  as the primary key of the respective relation. The relation $R$ is constrained by a single key dependency 
 $ B \rightarrow A $. The inclusion constraints are $B_{i,j}\subseteq B$, for each pair $i,j$ and
 $A\subseteq A_{i,j}$, also for each pair $i,j$.
 
 Consider a propositional formula $\Phi=\phi_1\wedge \phi_2 \wedge \ldots \phi_m$, where $\phi_i $ are clauses. 
 We assume that none of the clauses in $\Phi $ contains more than 3 literals, that
 each variable occurs at most 3 times in $\Phi $, and that the number of variables in $\Phi$
 is equal to the number $m$ of clauses in the formula. It is easy to prove that satisfiability is
 NP-complete even for formulae of this kind. For the formula $\Phi$ we built a database instance
 $ r_{\Phi}$: in the the relation $R$ we remember the formula $\Phi$: it
  consists of such pairs $(w,\phi)$ that $w$ is a literal, $\phi$ is a clause from $\Phi$
 and $w$ occurs in $\phi$. The definitions of the relations $R_{i,j}$ are a little bit  more complicated.
 The relation $R_{i,j}$ consists of $2m$ tuples 
 $(p_l,\phi_{s(i,j,l)})$, and $(\neg p_l,\phi_{s(i,j,l)})$, with $s$ still to be defined, will be a function 
 from $\{1,2,3\}\times \{1,2,3\} \times \{1,2, \ldots m\}$ to  $\{1,2, \ldots m\}$ and, more precisely,
  it is going to be a permutation
 of $\{1,2, \ldots m\}$ for every fixed pair $(i,j)$. 
 Define $s(i,j,l)$ as $n$ if $p_l$ (or $\neg p_l$) occurs in the clause 
 $\phi_{n+1}$ (where addition is modulo the  number of clauses $m$), if $p_l $ is the $i$th
 variable in this clause, and if it is $j$th occurrence of $p_l$ in $\Phi$. Now, for each $(i,j)$ 
 let $s(i,j,\_ ) $ be any permutation consistent with the above definition. 
 It follows directly from our construction that:

\begin{lemma}\label{pomocniczy}
For each clause $\phi_n$ from $\Phi$ and for each variable $p$ occurring in $\phi_n$ there
is a relation $R_{i,j}$ such that the tuples $(p,\phi_{n-1})$ and $(\neg p,\phi_{n-1})$ are
in $R_{i,j}$.
\end{lemma}

We want to  show   that $\emptyset$ is a repair of
$ r_{\Phi}$ with respect to $IC$ if and only if $\Phi $ is not satisfiable. 

The {\em only if} direction is simple. Assume that $\Phi$ is satisfiable and let $\rho $ be 
a satisfying assignment. In each tuple in each of the relations $R, R_{i,j}$ in $ r_{\Phi}$ 
the first argument  is always a literal. Let $ r'$ be a subset of $ r_{\Phi}$
consisting of such facts $R(w,\phi)$ or $R_{i,j}(w,\phi)$ that $\rho(w)=1$. The key constraints for 
$R_{i,j}$ are satisfied in $ r'$. The inclusion constraints $B_{i,j}\subseteq B $ are 
satisfied because, since $\rho$ was an  assignment satisfying $\Phi$,
$B=\{\phi_1,\phi_2,\ldots \phi_m\}$.
Also the inclusions $A\subseteq A_{i,j}$ hold. But the key dependency $B\rightarrow A$ does not need to
hold in $ r'$ (this is because there is possibly more than one literal $w$ in some clause such that
$\rho(w)=1$).
To construct a nonempty repair of $ r_{\Phi}$
take now $ r''$  built with the same relations $R_{i,j}$  as $ r'$ and with relation $R$
 being the result of selecting from the relation $R$ in $ r'$ exactly one tuple $(w,\phi)$ for each $\phi$.
 
The {\em if} direction is more complicated. If $ r'$ is a repair of $ r_{\Phi}$
then, in each of the relations $R_{i,j}$, for each clause $s(i,j,l)$ at most one of the tuples 
$(p_l,\phi_{s(i,j,l)})$  and $(\neg p_l,\phi_{s(i,j,l)})$ can be in $R_{i,j}$. This implies that at most
one of the literals $p_l,\neg p_l$ can be in $A_{i,j}$. But $A\subseteq A_{i,j}$ and, since 
$\Phi $ is not satisfiable, there must be a clause $\phi_l$ such that none of the literals 
from $\phi_l$ is in $A$. This means that $\phi_l$ is not in $B$. Consider the clause $\phi_{l+1}$.
By Lemma \ref{pomocniczy} for each  variable $p$ from  $\phi_{l+1}$
there
is a relation $R_{i,j}$ such that the tuples $(p,\phi_l)$ and $(\neg p,\phi_l)$ are
in $R_{i,j}$ in $ r_\Phi$.
But, by the inclusion constraints,
each of the $B_{i,j}$ should be a subset of $B$, so since  $\phi_l$ is not in $B$ in $ r'$
it is also not in any of the $B_{i,j}$ in $ r'$. While removing $\phi_l$ from $B_{i,j}$
we also delete the variables occurring in a tuple of $R_{i,j}$ together with $\phi_l$.
This means that for each variable $p$
from the clause $\phi_{l+1}$ there is a relation $R_{i,j}$ such that neither $p$ nor $\neg p$ is in
$A_{i,j}$. But $A$ is a subset of each of the $A_{i,j}$. This means that none of the literals
from $\phi_{l+1}$ can be in $A$. So $\phi_{l+1}$ cannot be in $B$! Now,  using this argument $m$ times 
we can remove all the tuples from the relations, thus proving that $  r'$ is empty.
\end{proof}
 
\subsection{Arbitrary FDs and INDs}\label{sec:general}

\begin{theorem}\label{t:general}
The repair checking problem for arbitrary FDs and INDs is co-NP-complete.
\end{theorem}
\begin{proof}
Co-NP-hardness was established earlier in this section.
The membership in co-NP follows from the definition of repair.
\end{proof}

\begin{theorem}
The consistent query answers problem for arbitrary FDs and INDs is $\Pi^p_2$-complete.
\end{theorem}
\begin{proof}
The membership in $\Pi^p_2$ follows from the definition of consistent query answer.
We show $\Pi^p_2$-hardness below.
Consider a quantified boolean formula $\phi $ of the form
\[\forall p_1,p_2,\ldots p_k \exists q_1, q_2,\ldots q_l \; \psi \] where 
$\psi$ is quantifier-free and equals to $\psi_1 \wedge \psi_2 \wedge \ldots \psi_m $,
where $\psi_i $ are clauses. 
We will construct a database instance $r_\phi$, over a schema with a single relation $R(A,B,C,D)$,
such that $R(a,a,\psi_1,a) $ is a consistent answer if and only if $\phi $ is true. 
The integrity constraints will be $A\rightarrow B$  and $C\subseteq D$.

There are 3 kinds of tuples in  $r_\phi$.
For each occurence of a literal in $\psi $ we have one tuple of the first kind
(we adopt the convention that $\psi_{m+1}$ is $\psi_1$):
\begin{itemize}
\item $ R(p_i,1,\psi_j, \psi_{j+1}) $  if $p_i$ occurs positively in $\psi_j$,
\item $ R(q_i,1,\psi_j, \psi_{j+1}) $  if $q_i$ occurs positively in $\psi_j$,
\item $ R(p_i,0,\psi_j, \psi_{j+1}) $  if $p_i$ occurs negatively in $\psi_j$,
\item $ R(q_i,0,\psi_j, \psi_{j+1}) $  if $q_i$ occurs negatively in $\psi_j$.
\end{itemize}

For each universally quantified variable $p_i$ we have two tuples of the second kind:
$R(p_i,1, a_i, a_i)$ and $R(p_i,0, a_i, a_i)$.
Finally, there is just one tuple of the third kind: 
$R(a,a,\psi_1,a)$.
\ignore{
The tuples of the first kind establish a correspondence between the repairs and the
assignments satisfying $\phi$. The tuples of the second kind generate all possible assignments
for the universally quantified variables. Finally, the tuple of the third kind is used to test
whether a complete assignment for the existentially quantified variables has been built.
}

Let us first show that if $\phi $ is false then $R(a,a,\psi_1,a)$ is not a consistent answer.
Let $\sigma $ be such a valuation of the variables $p_1,p_2, \ldots p_k$ that the formula
$\sigma(\phi)$
(with free variables $q_1,q_2,\ldots q_l$ is not satisfiable. It will be enough to show that the set
$s_\sigma $ of 
all the tuples from $ r_\phi$ which are of the form  $R(p_1,\sigma(p_i),a_i,a_i) $ is a repair.
The set $ s_\sigma $ is consistent. So if it is not a repair then another consistent subset $
s\supset  s_\sigma $ of  $ r_\phi$ 
must exist. Due to the FD  $ s$ does not contain any tuple of the second kind not being
already in  $ s_\sigma $. So, there must be some tuple of the first or the third kind in  $ s$. But that
means (due to the IND)
that for each $\psi_j$ there is either some tuple of the form $R(p_i,\sigma(p_i),\psi_j,\psi_{j+1}) $ in
$ s$ , or some tuple
of the form  $R(p_i,\varepsilon_i,\psi_j,\psi_{j+1}) $, where $\varepsilon_i\in\{0,1\}$. Due to the FD,
for each $q_i$ there can be 
at most one such $\varepsilon_i $. Define $\bar{\sigma}(q_i) = \varepsilon_i$. Then
$\bar{\sigma}(\sigma(\phi))=1$ which is 
impossible. 

For the opposite direction suppose that $\phi $ is true but  $R(a,a,\psi_1,a)$ is not a consistent
answer.
The last means that there exists a repair $ s$ of $ r_{\phi}$ such that no tuple of the form 
$R(\_,\_,\psi_1,\_) $ can be found in $s$. But this implies that there are no tuples of the
first kind in $ s$, and so $ s$  only consists of some tuples of the second kind. Due to the FD 
there exists a valuation $\sigma $ such that 
$ s$ 
consists of all the tuples of the second kind which are of the form $R(p_1,\sigma(p_i),a_i,a_i) $.
Since $\phi $ is true,
there exists a valuation  $\bar{\sigma}$ of variables $q_1,q_2, \ldots q_l$ such that
$\bar{\sigma}(\sigma(\phi))=1$.
But then the set  $ s'$ consisting of all the tuples from $ s$, $R(a,a,\psi_1,a)$, and
all the tuples of the first kind which are either of the form  $R(p_i,\sigma(p_i),\psi_j,\psi_{j+1}) $
or 
 $R(q_i,\bar{\sigma}(q_i),\psi_j,\psi_{j+1}) $ is consistent, which contradicts the assumption that $
s$ is a repair.
\end{proof}

\section{Related work}\label{sec:related}

We only briefly survey the related work here. A more comprehensive
discussion can be found in \cite{ArBeCh99,BeCh03}.

There are several similarities between our approach to consistency
handling  and those followed by the belief revision/update
community \cite{GaRo95}. Database repairs (Definition \ref{def:repair}) coincide
with revised models defined by Winslett in
\cite{Win88}. The treatment in \cite{Win88} is mainly propositional, but
a preliminary extension to first order knowledge bases can be
found in \cite{ChWi94}. Those papers concentrate on the
computation of the models of the revised theory, i.e., the repairs
in our case. Comparing our framework
with that of belief revision, we have an empty domain theory, one
model: the database instance, and a revision by a set of ICs.
The revision of a database instance by the ICs produces new
database instances, the repairs of the original database.
The complexity of belief revision (and the related problem of counterfactual
inference which corresponds to our computation of consistent query answers)
in the propositional case was exhaustively classified by Eiter and Gottlob \cite{EiGo92}.
Among the constraint classes considered in the current paper, only denial
constraints can be represented propositionally by grounding.
However, such grounding results in an unbounded update formula, which
prevents the transfer of any of the PTIME upper bounds from \cite{EiGo92}
into our framework. Similarly, their lower bounds require different kinds of
formulas from those that we use.

The need to accommodate violations of functional dependencies
is one of the main motivations for considering disjunctive
databases \cite{ImNaVa91,ChoSaa98:incom} and has led
to various proposals in the context of data integration
\cite{AgKeWiSa95,BaKrMiSu92,Dung96,LiMe96}.
There seems to be an intriguing connection between relation repairs w.r.t. FDs
and databases with disjunctive information \cite{ChoSaa98:incom}.
For example, the set of repairs of the relation {\em Person} from 
Example \ref{ex:expo} can be represented as a disjunctive database $D$
consisting of the formulas 
\[{\it Person\/}({\rm Brown},{\rm Amherst},{\rm 115\; Klein})
\vee {\it Person\/}({\rm Brown},{\rm Amherst},{\rm 120\; Maple})\]
and 
\[{\it Person\/}({\rm Green},{\rm Clarence},{\rm 4000\; Transit}).\]
Each repair corresponds to a minimal model of $D$ and vice versa.
We conjecture that the set of all repairs of an instance w.r.t. a set of FDs
can be represented as a disjunctive table (with rows that are disjunctions
of atoms with the same relation symbol).
The relationship in the other direction does not hold, as shown by the folowing example \cite{ABCHRS03}.
\begin{example}
The set of minimal models of the formula
\[(p(a_1,b_1)\vee p(a_2,b_2))\wedge p(a_3,b_3)\]
cannot be represented as a set of repairs of any set of FDs.\hfill $\Box$
\end{example}
Known tractable classes of first-order queries over disjunctive databases 
typically involve conjunctive queries and databases with restricted OR-objects
\cite{ImNaVa91,ImVMVa95}.
In some cases, like in Example \ref{ex:expo}, the set of all repairs can be represented 
as a table with OR-objects.
But in general this is not the case \cite{ABCHRS03}.
\begin{example}
Consider the following set of FDs $F=\{A\rightarrow B,A\rightarrow C\}$,
which is in BCNF.
The set of all repairs of the instance $\{(a_1,b_1,c_1),(a_1,b_2,c_2)\}$
cannot be represented as a table with OR-objects.\hfill $\Box$
\end{example}
The relationship in the other direction, from tables with OR-objects
to sets of repairs, also does not hold.
\begin{example}
Consider the following table with OR-objects:
\begin{center}
\begin{tabular}{|ll|}
\hline
OR(a,b) & c\\
a & OR(c,d)\\\hline
\end{tabular}
\end{center}
It does not represent the set of all repairs of any instance under
any set of FDs. \hfill $\Box$
\end{example}
In general, a correspondence between sets of repairs and tables with OR-objects
holds only in the very restricted case when the relation is binary,
say $R(A,B)$, and there is one FD $A\rightarrow B$.
The paper \cite{ImVMVa95} contains a complete classification of the
complexity of conjunctive queries for tables with OR-objects.
It is shown how the complexity depends on whether the tables satisfy
various schema-level criteria, governing the allowed occurrences
of OR-objects. Since there is no exact correspondence between tables
with OR-objects and sets of repairs of a given database instance,
the results of \cite{ImVMVa95} do not directly translate
to our framework, and vice versa. 

There are several proposals for language constructs specifying nondeterministic queries
that are related to our approach ({\em witness} \cite{AbHuVi95},
{\em choice} \cite{GiGrSaZa97,GiPe98,GrSaZa95}).
Essentially, the idea is to construct a maximal subset of a given relation that
satisfies a given set of functional dependencies. Since there is usually more than
one such subset, the approach yields nondeterministic queries in a natural way.
Clearly, maximal consistent subsets (choice models \cite{GiGrSaZa97}) correspond to repairs.
Datalog with choice \cite{GiGrSaZa97} is, in a sense, more general than our approach,
since it combines enforcing functional dependencies with inference using Datalog
rules. 
Answering queries in all choice models ($\forall G$-queries \cite{GrSaZa95}) corresponds to our notion of computation
of consistent query answers (Definition \ref{def:cqa}).
However, the former problem is shown to be co-NP-complete and no tractable cases are
identified. One of the sources of complexity in this case is the presence of Datalog
rules, absent from our approach.
Moreover, the procedure proposed in \cite{GrSaZa95}
runs in exponential time if there are exponentially many repairs, as in Example
\ref{ex:expo}. Also, only conjunctions of literals are considered as queries in \cite{GrSaZa95}.

A purely proof-theoretic notion of consistent query answer comes
from Bry  \cite{Bry97}.
This notion, described only in the propositional case, corresponds to 
evaluating queries after all the tuples involved in inconsistencies
have been eliminated.
The paper \cite{ArBeCh99} introduced the notions of repair and consistent query
answer used in the current research. It proposed computing consistent query answers through {\em query transformation}.
The papers \cite{ArBeCh01,ABCHRS03} studied the computation of consistent query
answers in the context of FDs and scalar aggregation queries.

Wijsen \cite{Wij03} studied the problem of consistent query answering 
in the context of universal constraints.
In contrast to Definition \ref{def:repair}, he considers repairs obtained
by modifying individual tuple components.
Notice that a modification of a tuple component
cannot be necessarily simulated as a  deletion followed
by an insertion, because this might not be minimal under
set inclusion. Wijsen proposes to represent all the repairs of an
instance 
using a single {\em trustable tableau}. From this tableau, answers
to conjunctive queries can be efficiently obtained. It is not clear,
however, what is the computational complexity of constructing the
tableau, or even whether the tableau is always of polynomial size.

Representing repairs as stable models of logic programs with disjunction and
classical negation has been proposed in \cite{ArBeCh00,GrGrZu01}.
Those papers consider computing consistent answers to first-order queries.
While the approach is very general, no tractable cases beyond those already
implicit in the results of \cite{ArBeCh99} are identified.
The semantics of referential integrity actions are captured using stable
models of logic programs with negation in \cite{LuMaLa97}.

It is interesting to contrast our results in Section \ref{sec:ind} with the classical results about the
implication problem for FDs and INDs \cite{AbHuVi95}. This problem is undecidable
in general but becomes decidable under suitable restrictions on INDs.
For instance, it is decidable in PTIME if the INDs are unary
and in EXPTIME if the INDs are acyclic. The problems discussed
in our paper are all in $\Pi^p_2$ (Section \ref{sec:ind}). The role the syntactic restrictions
play in this context is different.
The restriction to unary INDs is not helpful, c.f., Theorem \ref{t:general}.
The restriction to acyclic INDs makes the repair checking problem tractable (Theorem \ref{t:acyclic}) but not
so the problem of consistent query answers (Theorem \ref{t:acyclic:cqa}).

In \cite{MaRa92}, several classes of FDs and INDs were identified for which the implication
problem does not exhibit any interaction between the FDs and the INDs.
I.e., a set of constraints implies an FD (resp. an IND) iff the FDs (resp. the INDs) in this set
imply it.
Unfortunately, the syntactic restrictions on constraints that guarantee no interaction in the
above sense do not play a similar role in our context. It seems that the notion of maximality
present in the repair definition forces a  relationship between the FDs and the
INDs that is much tighter than the one implicit in the implication problem.

In \cite{MaMa90,MaRa92}, it is investigated what kind of relational schemas and integrity constraints
can result from mapping an Entity-Relationship schema (this is a common way of designing relational 
schemas). Acyclicity of INDs is a necessary requirement, thus repair checking is tractable in this case.
However, it turns out that the schema from Theorem
\ref{t:acyclic:cqa} could result from such a mapping. Thus, even restricting the relational schemas
to those that correspond to Entity-Relationship schemas does not guarantee the tractability of consistent
query answers.

\section{Conclusions and future work}\label{sec:concl}
In this paper we have investigated the computational complexity issues involved
in minimal-change   integrity maintenance using tuple deletions, in the presence of denial
constraints and inclusion dependencies. We have identified several
tractable cases and shown that generalizing them leads to intractability.

We envision several possible directions for future work. First, one
can consider various {\em preference orderings} on repairs. Such
orderings are often natural and may lead to further tractable
cases. Some preliminary work in this direction is reported in
\cite{GrGrZu01}.  Second, a natural scenario for applying the results
developed in this paper is {\em query rewriting} in the presence of
distributed data sources \cite{DuGeLe00,Hal01,Len02}. Recent work in this
area has started to address the issues involved in  data sources being
inconsistent \cite{BeChCoGu02,LeLeRo02}.  Finally, as XML is playing
an increased role in data integration \cite{PaVa99,LuPaVe00,DrHaWe01}, it would
be interesting and challenging to develop the appropriate notions of
repair and consistent query answer in the context of XML
databases. Recent integrity constraint proposals for XML include
\cite{BDFHT01,FaSi00,FaKuSi01}.

\newcommand{\etalchar}[1]{$^{#1}$}

 \end{document}